\documentclass[manuscript]{aastex}

\usepackage{natbib, aas_macros}
\usepackage{amsmath}
\usepackage{color}

\usepackage{ulem}

\definecolor{MyDarkBlue}{rgb}{0,0.08,0.45}
\definecolor{MyDarkRed}{rgb}{0.8,0.1,0.08}
\definecolor{Red}{rgb}{1.0,0.0,0.2}
\definecolor{Blue}{rgb}{0,0.08,0.95}
\definecolor{LightGrey}{rgb}{0.7,0.7,0.7}

\begin{document}
\title{Spatial inhomogeneity of kinetic and magnetic dissipations in thermal convection}
\author{H. Hotta}
\affil{Department of Physics, Graduate School of Science, Chiba
university, 1-33 Yayoi-cho, Inage-ku, Chiba, 263-8522, Japan
}
  \begin{abstract}
   We investigate the inhomogeneity of kinetic and magnetic dissipations in
   thermal convection using
   high-resolution calculations. In statistically steady turbulence, the
   injected and dissipated energies are balanced. This means that a large
   amount
   of energy is continuously converted to internal energy via 
   dissipation. As in thermal convection downflows are colder than upflows, 
   the inhomogeneity of the dissipation potentially changes the
   convection structure. Our investigation on the inhomogeneity of the
   dissipation shows the following. 1. More dissipation is seen
   around the bottom of the calculation domain, and this tendency is
   promoted with the magnetic field. 2. The dissipation in the downflow is much
   larger than that in the upflow. The dissipation in the downflow is more
   than 80\% of the total at maximum. This
   tendency is also promoted with the magnetic field. 3. Although
   2D probability density functions of the kinetic
   and
   magnetic dissipations versus the vertical velocity are similar, the
   kinetic and magnetic dissipations are not well correlated.
   Our result suggests that the spatial inhomogeneity
   of the dissipation is significant and should be considered when
    modeling a small-scale strong magnetic field generated with an
   efficient small-scale dynamo for low-resolution
   calculations.
  \end{abstract}

\clearpage
\section{Introduction}
Thermal convection is one of the most important processes for understanding
the differential rotation and the dynamo of the Sun. Because the rotation
has a crucial
role in introducing the anisotropy of the turbulence, it is important to
estimate the convection velocity in the solar convection zone. When the
convection velocity is fast (slow), the influence of the rotation is weak
(strong). Recent observational and theoretical investigations suggest
that the convection velocity in numerical calculations are
significantly faster than in the real Sun
\citep{2012PNAS..10911928H,2014ApJ...793...24L,2015ApJ...798...51H,2016ApJ...818...32F,2016ApJ...830L..15F}.
This is an important unsolved problem in solar physics. A possible
mechanism for suppressing the convection velocity is a magnetic
field. \cite{2015ApJ...803...42H} found that the small-scale dynamo is very
efficient in the solar convection zone, and the small-scale magnetic
field can reach the equipartition level of the kinetic energy. This
strong magnetic field acts like viscosity and reduces the convection
velocity by factor of 2 compared with the case without the magnetic
field. To date, suppression by the magnetic field alone could not resolve
the fast convection problem, but this could contribute to solving the problem. The
viscosity-like behavior of the magnetic field also suggests that
the real solar convection could be mimicked by an enhanced
viscosity instead of a strong magnetic field. As the strong
small-scale magnetic field is only achievable with a high resolution---
i.e., high Reynolds numbers,
a calculation that requires a huge numerical resource---it would be useful to explore
the possibility of mimicking the magnetic field by the viscosity to
reduce the numerical cost. Such an approach has been taken recently by
\cite{2016AdSpR..58.1475O}. They found that an increase in the thermal
Prandtl number $\mathrm{Pr}=\nu/\kappa$, where $\nu$ and $\kappa$ are
viscosity and thermal conductivity, respectively, decreases the convection
velocity. This result potentially indicates that a small-scale
strong magnetic field contributes to suppressing the convective velocity,
as the Lorentz force does not act as thermal conductivity
($\kappa$) but as viscosity ($\nu$).
\par
In this paper, we explore the overlooked physical processes by
only adding the viscosity to mimic the magnetic field.
Of course, there is an important difference between the forms of the
Maxwell
and viscous stress tensors. We disregard this difference
and focus on the dissipation. In a statistically steady turbulence, like
the solar convection, the energy injection and the viscous and diffusive
dissipations are balanced; i.e., all the injected energy must be
dissipated in the end. Thus, a significant amount of the energy is
continuously
dissipated in small-scale, where the viscosity and diffusivity are effective.
As there is no special location in isotropic
turbulence \citep[e.g.][]{2014ApJ...791...12B}, the location of the dissipation
is not important. In contrast, in thermal convection turbulence, the
structures of the up and down flows are different. Typically, the downflow
is colder than the upflow. When the kinetic and magnetic energies are
dissipated more in the downflow region, this can change the thermal
structure. In this study, we address this issue with high-resolution
hydrodynamic and magnetohydrodynamic calculations. The effect of the
magnetic field on the
dissipation is also investigated.
The strong magnetic field suggested in \cite{2015ApJ...803...42H} possibly
changes the character of the dissipation, and we need to consider
the difference in dissipation in low-resolution calculations when
the strong magnetic field is mimicked by enhanced viscosity \citep{2016AdSpR..58.1475O}.\par
In addition, we investigate the effect of the magnetic Prandtl number
($\mathrm{Pm}=\nu/\eta$, where $\eta$ is the magnetic diffusivity) on
the inhomogeneity of
dissipation.
A small magnetic Prandtl number ($\eta > \nu$) makes the scale
of the magnetic dissipation larger than that of the kinetic
dissipation and vice versa. As a result,
more (less) energy is dissipated through the magnetic dissipation with a
small (large) magnetic Prandtl number.
\cite{2011AN....332...51B,2014ApJ...791...12B} found
that the ratio of the dissipation ($\epsilon_\nu/\epsilon_\eta$, where
$\epsilon_\nu$ and $\epsilon_\eta$ are the kinetic and magnetic
dissipation) decreases with decreasing the magnetic Prandtl number.
In the solar convection zone, the magnetic Prandtl number is very small
\cite[$\mathrm{Pm}\sim10^{-5}-10^{-6}$][]{1996Sci...272.1286C,2005LRSP....2....1M}.
If the magnetic Prandtl number changes the dissipation character, this
should be considered even in high-resolution calculations when
$\mathrm{Pm}\sim 1$ is used.
\section{Model}
We solve the three-dimensional magnetohydrodynamic (MHD) equations in the
Cartesian geometry $(x,y,z)$. Here, we define the $x$-direction as the
gravity direction; the $y$- and $z$-directions are horizontal directions. The MHD
equations with the gravity and radiative heating and cooling are expressed
as:
\begin{eqnarray}
 \frac{\partial \rho}{\partial t} &=& -\nabla\cdot(\rho{\bf v}),\\
 \frac{\partial}{\partial t}(\rho{\bf v}) &=& -\nabla\cdot
  \left[
   \rho{\bf vv} + \left(p{\bf I} + \frac{B^2}{8\pi}{\bf I}-\frac{\bf BB}{4\pi}
		  \right) 
				  \right]- \rho g{\bf e}_x,\\
 \frac{\partial {\bf B}}{\partial t} &=& \nabla\times({\bf v}\times{\bf
  B}),\\
 \frac{\partial E_\mathrm{total}}{\partial t} &=& -\nabla\cdot
  \left[\left(
	 E_\mathrm{total} + p + \frac{B^2}{8\pi}
		      \right){\bf v}
  -\frac{({\bf v}\cdot{\bf B}){\bf B}}{4\pi}
				 \right]-\rho v_xg + \Gamma,\\
 E_\mathrm{total} &=& \frac{p}{\gamma-1} + \frac{1}{2}\rho v^2 +
  \frac{B^2}{8\pi}\label{energy},
\end{eqnarray}
where $\rho$, $p$, $g$, ${\bf v}$, ${\bf B}$, $E_\mathrm{total}$, and
$\Gamma$ are the density, the pressure, the gravitational acceleration,
the fluid velocity, the magnetic field, the total energy,
and the time-independent heating and cooling, respectively.
The ratio of the
heat capacities is $\gamma=5/3$.
We adopt the fourth-order space-centered derivative and the
Runge--Kutta method for time integration \citep{2005A&A...429..335V}. A stable
calculation is achieved with the artificial diffusivity suggested by
\cite{2009ApJ...691..640R} and \cite{2014ApJ...789..132R}. The details
of the artificial diffusivity are explained in the next section. In this
paper, we investigate the effect of a small-scale magnetic field on
the dissipation. The strong magnetic field is generated by an efficient
small-scale dynamo, which is only achievable by significantly reducing the viscosity
and the magnetic diffusivity. Thus, we do not include explicit
viscosity, magnetic diffusivity, and thermal conductivity. Only
the artificial diffusivity is included.
The initial conditions are the solution of the hydrostatic equilibrium
 $dp/dx =-\rho g$ with a constant gravitational acceleration. The
 temperature gradient is expressed as
 $\nabla=d\ln T/d\ln p$. Thus, the initial conditions are:
\begin{eqnarray}
 T_0 &=&T_\mathrm{b}\left[1-\nabla\frac{x}{H_\mathrm{b}}\right],\\
 p_0 &=&p_\mathrm{b}\left[1-\nabla\frac{x}{H_\mathrm{b}}\right]^{1/\nabla},\\
 \rho_0 &=&\rho_\mathrm{b}\left[1-\nabla\frac{x}{H_\mathrm{b}}\right]^{1/\nabla-1},
\end{eqnarray}
where $T_\mathrm{b}$, $p_\mathrm{b}$, $\rho_\mathrm{b}$, and
$H_\mathrm{b}=-(d\ln p/dx)^{-1}|_{x=0}=p_\mathrm{b}/\rho_\mathrm{b} g$ are
the temperature, the pressure, the density, and the pressure scale
height at $x=0$, which is the bottom of the calculation domain. In the
initial condition, the temperature gradient is set to the adiabatic
value $\nabla=(\gamma-1)/\gamma$ and becomes superadiabatic after
calculations start.
The calculation domain is $(0,0,0)<(x,y,z)/H_\mathrm{b}<(1.8,3.6,3.6)$.
We have numbers of grid points of 512, 1024, and 1024 in the $x$-, $y$-,
and $z$-directions, respectively.
We adopt the impenetrate stress-free boundary condition for
the velocity; i.e.,
$v_x=\partial v_y/\partial x=\partial v_z/\partial x=0$ at both the top
and bottom boundaries. At the top boundary, only the vertical magnetic
field is allowed, $\partial B_x/\partial x=B_y=B_z=0$, and the horizontal
magnetic field is allowed at the bottom boundary,
 $B_x=\partial B_y/\partial x=\partial B_z/\partial x=0$.
 The periodic boundary condition is adopted for all the variables in
 the horizontal direction.
$\Gamma$ is the time-independent cooling and heating around the top and
bottom boundaries expressed as:
\begin{eqnarray}
 F_\mathrm{r} &=&
  F_0
  \left(
  \exp\left[-\left(\frac{x-x_\mathrm{min}}{d_\mathrm{min}}\right)^2\right] +
  \exp\left[-\left(\frac{x-x_\mathrm{max}}{d_\mathrm{max}}\right)^2\right]
  \right)
  ,\\
 \Gamma &=& -\frac{dF_\mathrm{r}}{dx},
\end{eqnarray}
where $x_\mathrm{max}=1.8H_\mathrm{b}$ and $x_\mathrm{min}=0$ are the
locations of the top and bottom boundaries, respectively.
We set $d_\mathrm{max}=0.2H_\mathrm{b}$ and
$d_\mathrm{min}=0.4H_\mathrm{b}$ for the width of the heating and the
cooling, respectively.
$F_0=10^{-4}\rho_\mathrm{b}c_\mathrm{b}^3$ is the energy flux through the
calculation domain, where
$c_\mathrm{b}=\sqrt{\gamma p_\mathrm{b}/\rho_\mathrm{b}}$ is the speed
of sound at $x=0$.
The density contrast is $\rho_0(x_\mathrm{min})/\rho_0(x_\mathrm{max})\sim6.7$
in this setting.\par
Although the calculations in this study are toy models, we implicitly
assume that the computational domain extends from the base of the
convection zone to somewhere upper convection zone. We note that
the energy flux in this study
($F_0=10^{-4}\rho_\mathrm{b}c_\mathrm{b}^3$) is much larger than the
real solar value ($F_0\sim10^{-11}\rho_\mathrm{b}c_\mathrm{b}^3$).
\cite{2017ApJ...843...52K}, however, shows that the convection
property in the convection zone, is not influenced by the value of the
energy flux when the convection velocity is normalized with a typical
convection velocity $v_\mathrm{c}=(F_0/\rho_\mathrm{b})^{1/3}$.
\par

\subsection{Artificial diffusivity}

We adopt the artificial diffusivity (hyperdiffusivity) developed by
\cite{2009ApJ...691..640R} and \cite{2014ApJ...789..132R}.
The diffusion-like equation is applied for all the variables
($\rho$, ${\bf v}$, ${\bf B}$ and $e_\mathrm{in}$, where
$e_\mathrm{in}=p/(\gamma-1)$ is the internal energy):
\begin{eqnarray}
 \frac{u_i^{n+1}-u_i^{n}}{\Delta t} = -\frac{f_{i+1/2}-f_{i-1/2}}{\Delta
  x},
\end{eqnarray}
where the diffusive flux is expressed as:
\begin{eqnarray}
 f_{i+1/2} = -\frac{1}{2}c_{i+1}\Delta x
  \Phi(u_\mathrm{r}-u_\mathrm{l},u_{i+1}-u_i)(u_\mathrm{r}-u_\mathrm{l}).
\end{eqnarray}
The variables at the left and right sides of the cell surface are defined
as:
\begin{eqnarray}
 u_\mathrm{l} &=& u_i + 0.5\Delta u_i, \\
 u_\mathrm{r} &=& u_{i+1} - 0.5\Delta u_{i+1}.
\end{eqnarray}
The difference in the variable in the cell is defined with the
monotonized central minmod limiter as:
\begin{eqnarray}
 \Delta u_i = \mathrm{minmod}
  \left[
   \frac{u_{i+1}-u_{i-1}}{2},2(u_{i+1}-u_i),2(u_i-u_{i-1})
  \right]
\end{eqnarray}
The limiter function is defined with:
\begin{eqnarray}
 \Phi = \mathrm{max}
  \left[
   0,1+h\left(
	 \frac{u_\mathrm{r}-u_\mathrm{l}}{u_{i+1}-u_i} - 1
	\right)
  \right],\ \ &&(u_\mathrm{r}-u_\mathrm{l})(u_{i+1}-u_i) > 0\\
 \Phi = 0
  ,\ \ &&(u_\mathrm{r}-u_\mathrm{l})(u_{i+1}-u_i) \le 0,
\end{eqnarray}
where $h$ is a free parameter for the hyperdiffusivity. When we adopt
$h=0$, the diffusive flux is identical to the second-order local Lax--Friedrichs
scheme. A less diffusive calculation is achieved with larger $h$. In this
work, $h=1$ and $h=2$ are adopted for the density $\rho$ and the
internal energy $e_\mathrm{in}$, respectively.
\par
\begin{figure}[htbp]
 \centering
 \includegraphics[width=12cm]{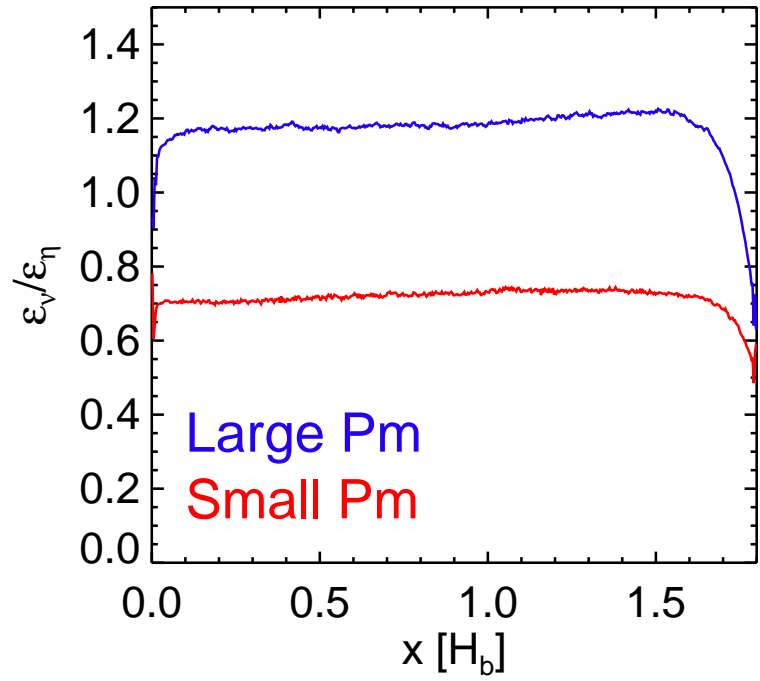}
 \caption{
 The ratio of the kinetic and magnetic dissipations
 $\epsilon_\nu/\epsilon_\eta$ is shown. Blue and red lines show the results
 in the Large Pm and Small Pm phases, respectively.
 \label{prandtl}}
\end{figure}

In this paper, the calculation includes three phases named Hydro, Large Pm,
and Small Pm. 
We start the calculation without the magnetic field
(phase Hydro). In this phase, $h=2$ is adopted for the velocity.
The calculation continues until the thermal convection reaches
a statistically steady state. Then, a weak horizontal seed magnetic field is
added (phase Large Pm). In this phase, $h=2$ is adopted for both the
velocity and the magnetic field.
When we adopt the same values of $h$ for the velocity and the
magnetic field in this setting, the magnetic dissipation
$\epsilon_\eta$ is slightly smaller than the kinetic dissipation
$\epsilon_\nu$ (Fig. \ref{prandtl}). Thus, the second phase is named
Large Pm.
Here we assume $\epsilon_\nu/\epsilon_\eta\sim \mathrm{Pm}$ \citep{2014ApJ...791...12B}.
When the small-scale dynamo reaches a statistically steady
state, the $h$ for the velocity increases to 500; i.e., the diffusivity for the velocity
decreases (phase Small Pm).

\par
The kinetic ($\epsilon_\nu$) and magnetic ($\epsilon_\eta$) dissipations at
a grid point i are estimated with:
\begin{eqnarray}
 \epsilon_\nu &=& -\frac{1}{2}\sum^3_{m=1}
  \left(
    \rho_{i-1/2}f^m_{i-1/2}(v^m)\frac{v^m_{i  }-v^m_{i-1}}{\Delta x}
  + \rho_{i+1/2}f^m_{i+1/2}(v^m)\frac{v^m_{i+1}-v^m_{i  }}{\Delta x}
  \right),\\
 \epsilon_\eta &=& -\frac{1}{8\pi}\sum^3_{m=1}
  \left(
    f^m_{i-1/2}(B^m)\frac{B^m_{i  }-B^m_{i-1}}{\Delta x}
  + f^m_{i+1/2}(B^m)\frac{B^m_{i+1}-B^m_{i  }}{\Delta x}
  \right),
\end{eqnarray}
respectively, where $m$ expresses three directions $x$, $y$, $z$ and
$\rho_{i+1/2}=(\rho_{i+1}+\rho_{i})/2$.
These dissipations are the loss of the kinetic ($\epsilon_\nu$) and
magnetic ($\epsilon_\eta$) energies through the artificial viscosity.
This is the mechanism to transform the kinetic and magnetic energies to
the internal energy in the calculations.
We note that because we use the total energy
$E_\mathrm{total}$ for the energy equation (\ref{energy}),
$\epsilon_\nu$ and $\epsilon_\eta$ are not included in the equation.
Any hyperdiffusivity becomes effective at a strong shear and current.
As we expect that
the hyperdiffusivity possibly overestimates the spatial inhomogeneity of the
dissipations, pseudo-dissipations $D_\nu$ and $D_\eta$ that are related to explicit
viscosity and magnetic diffusivity are additionally estimated as:
\begin{eqnarray}
D_\nu &=& \rho \sum_{i,k}\frac{\partial v_i}{\partial x_k}
\left[
\frac{\partial v_i}{\partial x_k} +
\frac{\partial v_k}{\partial x_i} -
\frac{2}{3}\delta_{ik}\nabla\cdot {\bf v}
	  \right],\\
 D_\eta &=&
  \frac{1}{4\pi}|\nabla\times {\bf B}|^2.
\end{eqnarray}
When the equation of motion has an explicit kinetic viscosity as
\begin{eqnarray}
 \frac{\partial }{\partial t}(\rho {\bf v}) = [...] - \nabla\cdot{\bf \tau},
\end{eqnarray}
where
\begin{eqnarray}
 \tau_{ij} &=& -2\rho \nu
  \left[
   e_{ij} - \frac{1}{3}(\nabla\cdot{\bf v})\delta_{ij}
	    \right],\\
 e_{ij} &=& \frac{\partial v_i}{\partial x_k},
\end{eqnarray}
the kinetic energy loss rate is expressed with the kinetic
pseudo-dissipation $D_\nu$ as $\nu D_\nu =
\rho\sum_{i,k}e_{ik}\tau_{ij}$. With the explicit magnetic diffusivity
as
\begin{eqnarray}
\frac{\partial {\bf B}}{\partial t} = [...] - \nabla\times
\left(\eta \nabla\times{\bf B}\right),
\end{eqnarray}
the magnetic
energy loss is expressed with the magnetic pseudo-dissipation $D_\eta$
as $\eta D_\eta=\eta|\nabla\times{\bf B}|^2/(4\pi)$.
Thus the pseudo-dissipations $D_\nu$ and $D_\eta$ could be the proxies
of the explicit viscosity and magnetic diffusivity.
The strong shear and current are 
selectively dissipated by the hyper diffusivity. The pseudo-dissipations likely underestimate the
inhomogeneity compared with reality, because we estimate these
just after the artificial viscosity is exerted; i.e., the structure
is smoothed. Thus, the realistic dissipation
inhomogeneity could be somewhere between the dissipation and the
pseudo-dissipation. In the following discussion, we show both the dissipations
($\epsilon_\nu$ and $\epsilon_\eta$) and the pseudo-dissipations ($D_\nu$
and $D_\eta$). We note that the ratios between the dissipations to the pseudo-
dissipations can be the effective viscosity ($\nu_\mathrm{eff}=\epsilon_\nu/D_\nu$)
and magnetic diffusivity
($\eta_\mathrm{eff}=\epsilon_\eta/D_\eta$). The above issues are also
discussed in \cite{2017ApJ...834...10R}.

\section{Result}

 \begin{figure}[htbp]
  \centering
  \includegraphics[width=12cm]{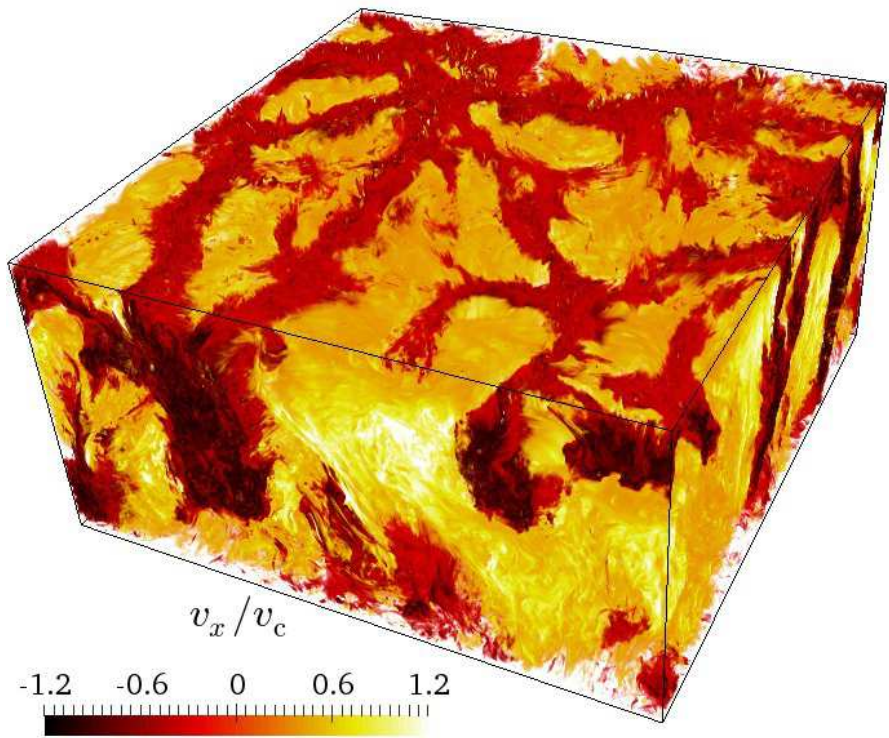}
  \caption{Volume rendered data of the vertical velocity ($v_x$) in the
  phase Large Pm.
  \label{volu}}
 \end{figure}

 Fig. \ref{volu} shows the volume-rendered data of the vertical velocity
 $v_x/v_\mathrm{c}$ in the phase Large Pm.
 The typical compressible thermal convection
 pattern---i.e., thin downflow surrounded by broad upflow--is observed. We
 achieve an efficient small-scale dynamo that suppresses the small-scale
 flows \citep{2015ApJ...803...42H,2016Sci....351..1427}.\par
  \begin{figure}[htbp]
  \centering
  \includegraphics[width=12cm]{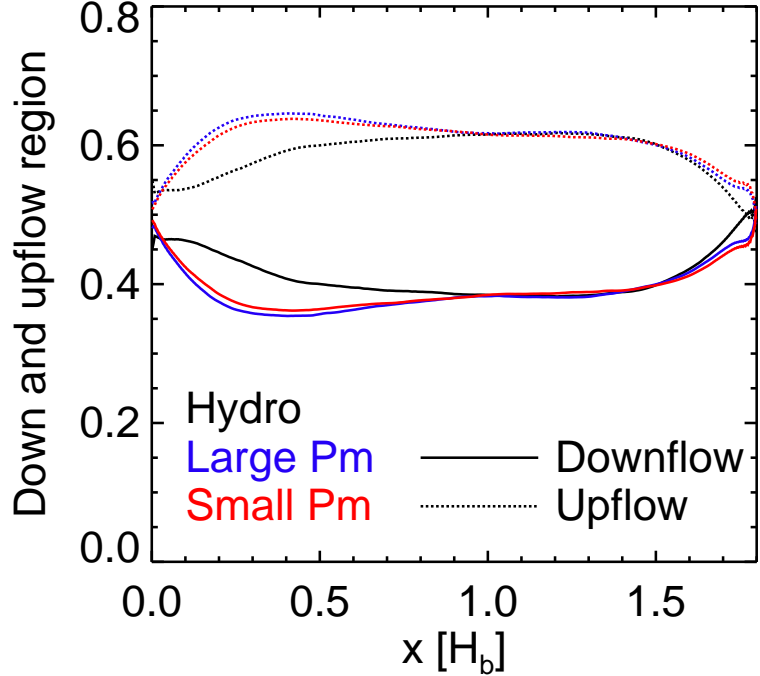}
  \caption{Dashed and solid lines show the fractions of the upflow and
   downflow regions, respectively.
   The results from the phases Hydro (black), Large Pm (blue), and Small
   Pm (red) are shown.
  \label{area}}
 \end{figure}

Fig. \ref{area} shows the fractions of the upflow (dashed) and downflow
(solid) regions. The results in the phases
Hydro (black), Large Pm (blue), and Small Pm (red) are shown.
About 60\% of the horizontal area is occupied by the broad upflow. This
feature is similar to the calculation with higher density
contrast \citep[$\sim600$:][]{2014ApJ...786...24H}. The results with the
magnetic field show a larger fraction of the upflow in the bottom half of
the computational domain.
Without the magnetic field, small-scale turbulent flow mixes the upflows
and downflows, and the region of the upflows decreases smoothly
toward the bottom of the calculation domain.
The strong magnetic field generated by the
small-scale dynamo suppresses the small-scale flow. Thus, the region of
the upflow increases because of the effect of stratification; i.e., the
downflow becomes thinner with higher gas pressure around the bottom
calculation domain.
No significant difference is observed between results with
different magnetic Prandtl numbers (blue and red lines).\par
   \begin{figure}[htbp]
  \centering
  \includegraphics[width=14cm]{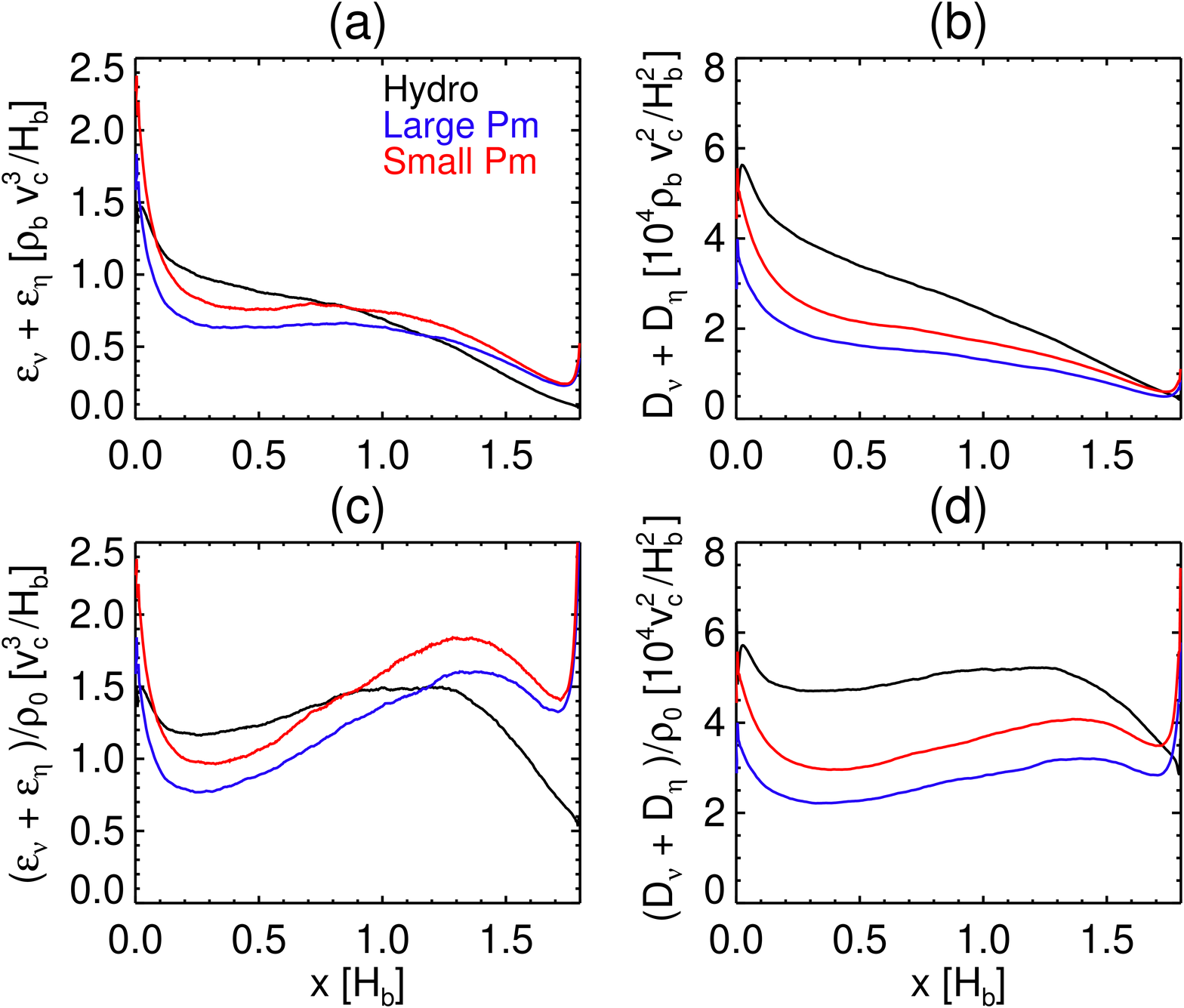}
    \caption{
    Horizontally averaged dissipations are shown.
    The results in the phases
    Hydro (black), Large Pm (blue), and Small Pm (red) are shown.
    Panels a and b show the total
    dissipation ($\epsilon_\nu+\epsilon_\eta$) and the total pseudo-dissipation
    ($D_\nu + D_\eta$), respectively. Panels c and d show the quotients of
    the panels a and b divided by the background density ($\rho_0$),
    respectively.
  \label{total}}
 \end{figure}

 Fig. \ref{total} shows the total dissipation
 (panel a: $\epsilon_\nu+\epsilon_\eta$) and
 the total pseudo-dissipation (panel b: $D_\nu+D_\eta$). For a comparison,
 panels c and d show quotients of the panels a and b divided by the
 background density ($\rho_0$), respectively.
 The results in the phases Hydro (black), Large Pm (blue), and Small Pm
 (red) are shown.
 The total dissipation density is estimated with vertical integration of the
 panel a as
 \begin{eqnarray}
  \int_{x_\mathrm{min}}^{x_\mathrm{max}}(\epsilon_\nu + \epsilon_\eta) dx
   \sim \rho_\mathrm{b} v_\mathrm{c}^3,
 \end{eqnarray}
 This is almost same value as the energy flux imposed from the bottom
 boundary. We note that the dissipated kinetic and magnetic energies to
 the internal energy can return back to the kinetic energy through the
 pressure and buoyancy works again. Thus the total dissipation density
 does not mean the increase rate of the internal energy.
 As expected, comparison between panels a and b shows
 that the inhomogeneity increased in the estimation of
 $\epsilon_\nu+\epsilon_\eta$; i.e., panel a shows steeper increase
 of the dissipation at the bottom of the computational domain than
 panel b.
 The increase in dissipation toward the bottom becomes larger when the magnetic
 field is included. \cite{2015ApJ...803...42H} shows that the downward Poynting
 flux efficiently transports magnetic energy to the bottom of the
 computational domain, and the small-scale dynamo is most effective
 around the bottom. Thus, a small-scale strong magnetic field is
 accumulated around the bottom. This magnetic field dissipates and
 contributes to the increase in dissipations around the bottom when
the magnetic field is included. This increase is also seen even in the
 total pseudo-dissipation (panel b: $D_\nu+D_\eta$), which likely reduces
 the space inhomogeneity; i.e., the increase in dissipation at the
 bottom is alleviated. In addition, this feature is seen even when it is divided by
 the background density (panels c and d).
 \cite{2017ApJ...843...52K}
 recently found that the small-scale dynamo
 with the artificial wall boundary at the bottom
 is not very much different from that with the radiation zone using the
 realistic solar parameter, since the radiation zone is significantly
 stiff in terms of the thermal convection. In other word, the artificial
 wall at the bottom nicely mimics the realistic solar radiation zone.
 Thus we argue that the
 increase of the dissipation toward the bottom boundary is not just an artifact.
 
\begin{figure}[htbp]
  \centering
  \includegraphics[width=14cm]{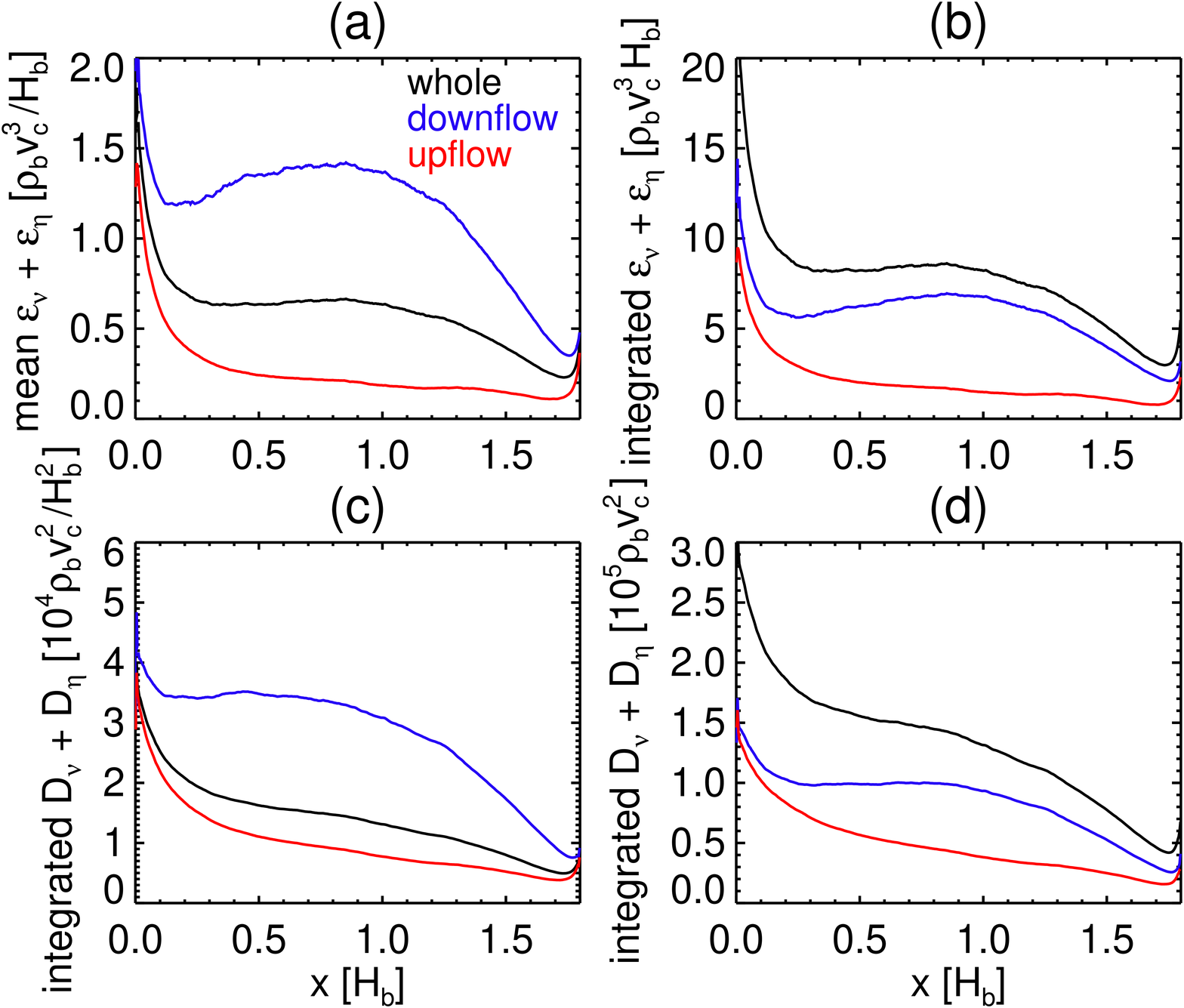}
    \caption{ Comparisons of the dissipations in the upflow and downflow
     regions are shown. The dissipations
     for the whole (black), downflow (blue), and upflow (red) regions are shown. Panels a and
     b show the mean and integrated dissipations
     ($\epsilon_\nu+\epsilon_\eta$), respectively. Panels c and d show
     the mean and integrated pseudo-dissipations ($D_\nu+D_\eta$), respectively.
  \label{updown_flow_hu}}
\end{figure}

Fig. \ref{updown_flow_hu} shows the dissipations at the upflow and downflow regions
separately in the Large Pm phase.
The results for the whole (black), downflow (blue), and upflow (red) regions
are shown.
In this paper, we show both the mean and integrated dissipation for
comparison.
The integrated dissipation at the downflow region is
calculated by integration of the dissipation over the downflow
region. Then, the integrated dissipation is divided by the area of the
downflow region, and the mean dissipation at the downflow region is calculated. The
same routine is used for the mean and integrated dissipation at the upflow region.
Panel a
shows the horizontally averaged (mean) dissipation. The figure shows that
the mean dissipations
in the downflow region (blue line) are much larger than those in the
upflow region (red line). As the downflow region is smaller than the
upflow region, it is useful to see the integrated dissipations in these
regions. Panel b shows the integrated dissipation over the corresponding
area. Even after integration, much more dissipation is observed in the
downflow region (Fig. \ref{updown_flow_hu}b).
This feature is observed even for the pseudo-dissipations, where inhomogeneity is
probably suppressed (Fig. \ref{updown_flow_hu}c and d). Thus, we
conclude that the dissipation is dominantly located in the downflow region.

 \begin{figure}[htbp]
  \centering
  \includegraphics[width=14cm]{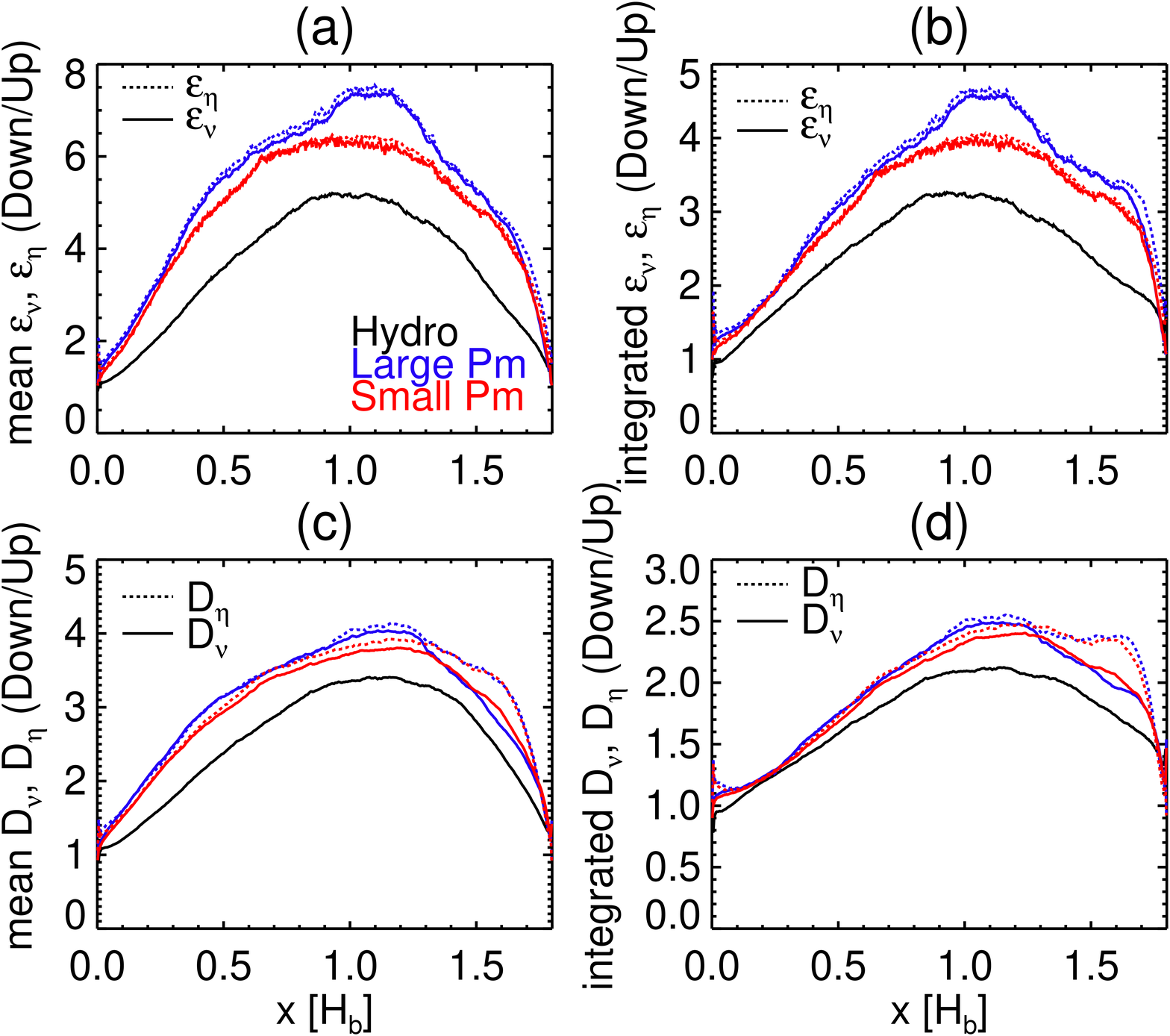}
       \caption{Ratios of dissipations in downflow to upflow regions
in the 
  Hydro (black), Large Pm (blue), and Small Pm (red) phases. In panels a (mean) and b
  (integrated), the solid and dashed lines show the kinetic
  ($\epsilon_\nu$) and magnetic ($\epsilon_\eta$) dissipations,
  respectively. In panels c (mean) and d (integrated), the solid and
  dashed lines show the kinetic ($D_\nu$) and magnetic ($D_\eta$) pseudo-
  dissipations, respectively.
  \label{updown_flow_ratio}}
 \end{figure}

Fig. \ref{updown_flow_ratio} shows the ratio of the dissipation in the
downflow region to that in the upflow region.
Panels a and b show the mean and integrated dissipation
($\epsilon_\nu$ and $\epsilon_\eta$), and panel c and d show the mean and
integrated pseudo-dissipations ($D_\nu$ and $D_\eta$), respectively.
The kinetic (solid:
$\epsilon_\nu$ and $D_\nu$) and
magnetic (dashed: $\epsilon_\eta$ and $D_\eta$) dissipations are shown
separately for the phases
Hydro (black), Large Pm (blue), and Small Pm (red). The kinetic
($\epsilon_\nu$) and magnetic ($\epsilon_\eta$) dissipations show almost the
same inhomogeneity between the upflow and downflow regions; i.e.,
the dashed and solid lines almost completely overlap.
We note that this does not mean $\epsilon_\nu=\epsilon_\eta$ and 
merely indicates that the ratios between upflow and downflow are the same.
All the results show that the largest and smallest inhomogeneities are 
shown in the Large Pm and Hydro phases, respectively. This means that
the magnetic field increases the dissipation in the downflow region.
The decrease in the Prandtl number reduces dissipation in the downflow
region. As we change the kinetic viscosity to change the
Prandtl number in this study, this means that the magnetic diffusivity stays the same,
and the pseudo-magnetic dissipation ($D_\eta$) is
almost identical between the Small Pm and Large Pm phases.
\par
 
 \begin{figure}[htbp]
  \centering
  \includegraphics[width=14cm]{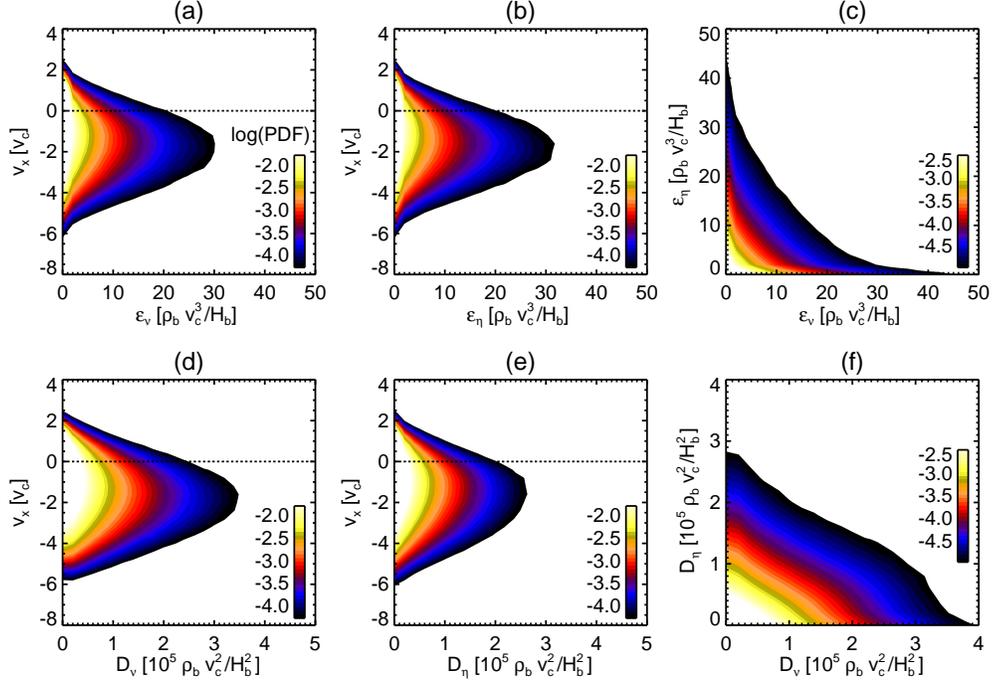}
       \caption{Two-dimensional (2D) probability density functions (PDFs) of
  the dissipations ($\epsilon_\nu$, $\epsilon_\eta$, $D_\nu$, and $D_\eta$)
  and the vertical velocity ($v_x$).
  Each panel shows a PDF of (a)
  $\epsilon_\nu$ and $v_x$, (b) $\epsilon_\eta$ and $v_x$, (c)
  $\epsilon_\nu$ and $\epsilon_\eta$, (d) $D_\nu$ and $v_x$, (e) $D_\eta$
  and $v_x$, and (f) $D_\nu$ and $D_\eta$.
  PDFs are calculated with the result in the Large Pm phase at
  $x=0.9H_\mathrm{b}$.
  A PDF of variables $a$ and $b$ ($F_{ab}$) are normalized to
  satisfy a relation $\int\int F_{ab}dadb=1$
  \label{pdf}}
 \end{figure}

 Fig. \ref{pdf} shows 2D probability density functions (PDFs).
 The panels show the PDF of (a)
 $\epsilon_\nu$ and $v_x$, (b) $\epsilon_\eta$ and $v_x$, (c)
 $\epsilon_\nu$ and $\epsilon_\eta$, (d) $D_\nu$ and $v_x$, (e) $D_\eta$
 and $v_x$, and (f) $D_\nu$ and $D_\eta$. Panels a, b, d, and e show similar
 behavior. The main dissipation occurs around $v_x\sim -2v_\mathrm{c}$, and these
 distributions are almost symmetrical with respect to the $v_x=-2v_\mathrm{c}$
 axis. Although no significant difference can be seen between the
 distribution of the kinetic and magnetic dissipations, panels c and f show
 that the correlation of the kinetic and magnetic dissipation is not large.
 When there is strong kinetic dissipation ($\epsilon_\nu$ or
 $D_\nu$), strong magnetic dissipation ($\epsilon_\eta$ or $D_\eta$) is
 hardly seen, and vice versa.

 \section{Summary and Discussion}
\begin{figure}[htbp]
 \centering
 \includegraphics[width=14cm]{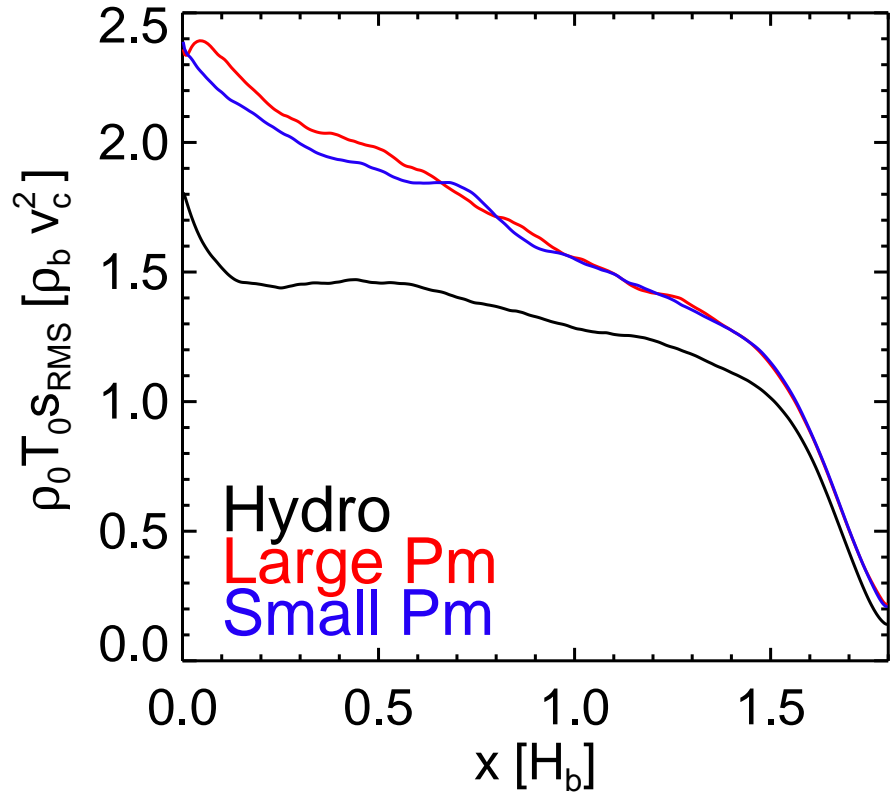}
 \caption {$\rho_0T_0 s_\mathrm{RMS}$ shown for the
 Hydro (black), Large Pm (blue), and Small Pm (red) phases.
 \label{serms}}
\end{figure}

 We perform high-resolution simulations to investigate the
 inhomogeneity of the kinetic and magnetic dissipations. Our main conclusions
 are as follows.
 \begin{enumerate}
  \item More dissipation is observed around the bottom of the calculation domain
	than that at the middle. This tendency is
	promoted with a magnetic field with an efficient small-scale
	dynamo around the bottom of the calculation domain.
  \item Dissipation in the downflow region is dominant. The ratio
	(downflow/upflow) is increased with the magnetic field, and a small
	magnetic Prandtl number produces a small ratio.
  \item Although 2D probability density functions show a similar
	distribution between kinetic and magnetic dissipation, the
	precise location of the strong dissipation is different; i.e.,
	the strong kinetic dissipation does not occur at the location
	where the strong magnetic dissipation occurs, and vice versa.
 \end{enumerate}
 Fig. \ref{serms} shows the root-mean-square (RMS) entropy
 perturbation. The results in the
 Hydro (black), Large Pm (blue), and Small Pm (red) phases are shown. The
 values are typically
 $\rho_0 T_0 s_\mathrm{RMS}\sim 2\rho_\mathrm{b}v_\mathrm{c}^2$ in the
 middle of the calculation domain. The
 difference in dissipation between upflow and downflow is typically
 $\Delta \epsilon \sim \rho_\mathrm{b}v_\mathrm{c}^3/H_\mathrm{b}$ (see
 panel a of Fig. \ref{updown_flow_hu}). This suggests that the difference in
 dissipation is large enough to diminish the thermal structure in the
 time scale of
 $\rho_0T_0s_\mathrm{RMS}/\Delta
 \epsilon\sim2H_\mathrm{b}/v_\mathrm{c}$, which is twice the turnover
 time of the convection cell. Although the thermal structure is not
 diminished because of continuous generation by the thermal convection
 itself, the difference in convection
 should not be ignored and should have a significant role in determining the
 thermal structure.
 The kinetic and magnetic energies are
 dissipated in the downflow with a magnetic field,
 indicating that the inhomogeneity of the 
 dissipation tends to decrease the RMS entropy and
 temperature perturbation. Fig. \ref{serms}, however, shows an increase
 in entropy perturbation with the magnetic
 field. \cite{2015ApJ...803...42H} show that a magnetic field
 suppresses the mixing between the downflow and upflow, and increases the
 entropy perturbation. As many mechanisms are involved in determining
 the thermal structure, the change in dissipation cannot directly
 change it. When we mimic a magnetic field with large viscosity in low-
 resolution calculations, we need to keep in mind the fact that the
 change in dissipation by a strong magnetic field may not be
 reproduced only with the enhanced viscosity.

\acknowledgements
The author would like thank anonymous referee for his/her great
suggestions which improve the manuscript.
The author is grateful to J. Warnakke and H. Iijima for helpful comments on the
manuscript. The author wants to thank T. Hanawa for stimulus discussions
in a seminar.
The results are obtained by using the K computer at the RIKEN Advanced
Institute for Computational Science (Proposal number hp170239,
hp170012, hp160026, hp160252,
ra000008). This work was supported by MEXT/JSPS KAKENHI Grant Number
JP16K17655, JP16H01169.
This research was supported by MEXT as ``Exploratory Challenge on Post-K
computer'' (Elucidation of the Birth of Exoplanets [Second Earth] and
the Environmental Variations of Planets in the Solar System).
This work is supported by Strategic International Research Exchange
Promotion Program in NINS (UFEX402).

\end{document}